\documentclass[manuscript,screen]{acmart}

\usepackage{caption,subcaption,booktabs}
\usepackage{graphicx}
\usepackage{enumitem}
\usepackage{footnote}
\usepackage{array} 
\usepackage{multirow}

\newcolumntype{C}[1]{>{\centering\arraybackslash}p{#1}}

\AtBeginDocument{%
  \providecommand\BibTeX{{%
    \normalfont B\kern-0.5em{\scshape i\kern-0.25em b}\kern-0.8em\TeX}}}

\setcopyright{rightsretained}
\copyrightyear{2025}
\acmYear{2025}
\acmDOI{XXXXXXX.XXXXXXX}

\acmConference[-]{-}{}{}
\acmBooktitle{-} 
\acmISBN{-}

\begin{document}

\newcommand{\Vis}{Visionary Tuning}
\newcommand{\Sys}{\textsc{VisionForge}}
\title[LLM Self-Playing and Self-Improving as a Design Tool]{Model Behavior Specification by Leveraging LLM Self-Playing and Self-Improving}


\author{Soya Park}
\affiliation{%
  \institution{Emory University}
  \country{USA}}
\email{soya.park@emory.edu}

\author{J.D. Zamfirescu-Pereira}
\affiliation{%
  \institution{UC Berkeley}
  \country{USA}
}

\author{Chinmay Kulkarni}
\affiliation{%
  \institution{Google \& Emory University}
  \country{USA}
}

\newcommand{\SP}[1]{\textcolor{purple}{\textbf{*Soya*}: #1}}
\newcommand{\jd}[1]{\textcolor{orange}{\textbf{*J.D.*}: #1}}
\newcommand{\ck}[1]{\textcolor{olive}{\textbf{*Chinmay*}: #1}}

\renewcommand{\shortauthors}{Soya Park, J.D. Zamfirescu-Pereira and Chinmay Kulkarni}

\begin{abstract}

Training AI models is challenging, particularly when crafting behavior instructions. Traditional methods rely on machines (supervised learning) or manual pattern discovery, which results in not interpretable models or time sink. While Large Language Models (LLMs) simplify instruction writing through natural language, articulating intended model behavior still remains difficult.

We introduce \textit{\Vis{}}, a human-in-the-loop self-playing followed by automatic self-refinement to improve behavior specification. Our system helps users clarify desired behavior through self-playing and generates prompts through self-improving, Our first evaluation involves user study conducted on a system implementation of \Vis{} within the context of chatbot behavior. Our system self-play itself by simulating user interactions to identify patterns and create effective prompts based on the pattern. In a within-subject study (N=12), participants pinpointed more patterns through self-playing and crafted better prompts. Surprisingly, users felt more or less success level in specifying the model behavior. Follow-up crowd studies (N=60) confirmed that the chatbot adhered to instructions without sacrificing quality. Our second evaluation is a case study on a real-world implementation using a movie rating dataset with \Vis{}, demonstrating its effectiveness and robustness in modeling a critic's preferences across the spectrum of low to highly rated movies.

Together, these results suggest how AI improves the design process of interactive AI systems. Furthermore, they suggest how the benefits of these tools may be non-obvious to end-users. We reflect on these findings and suggest future directions.   

\end{abstract}




\begin{teaserfigure}
  \includegraphics[width=\textwidth]{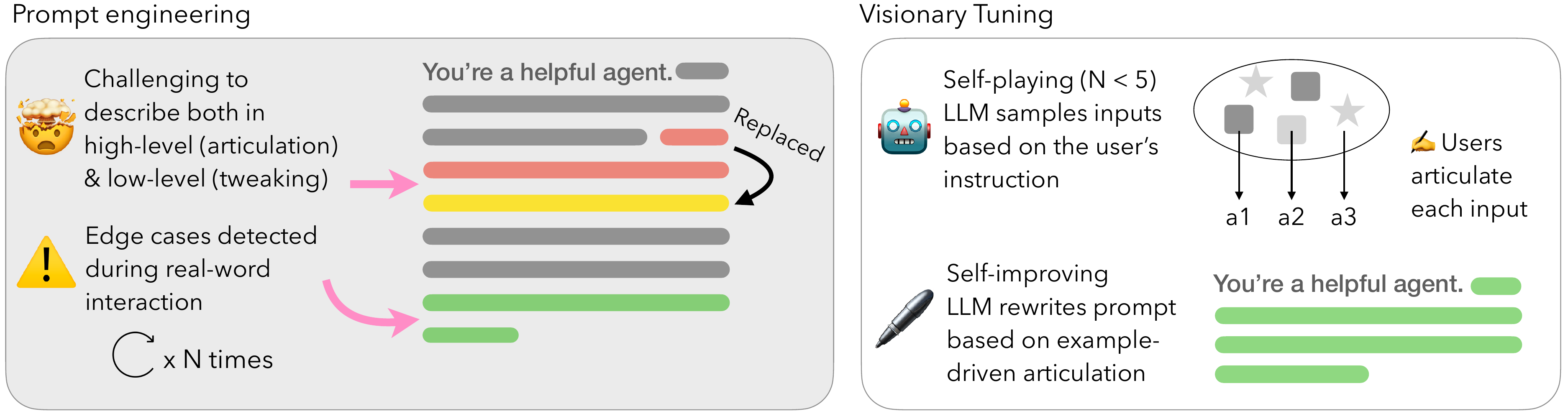}
  \caption{\Vis{} allows developers to craft model instructions by letting them focus on the high-level behavior articulation through two components: \emph{Self-playing phase}: a target model interacts with itself under the monitoring and supervision of the model developer. In this phase, the goal is to surface diverse criteria to probe users' training intentions. \emph{Self-improving phase}: Based on the user feedback, we use LLM to articulate instructions for the target model.}
  \label{fig:teaser}
\end{teaserfigure}

\received{20 February 2007}
\received[revised]{12 March 2009}
\received[accepted]{5 June 2009}

\maketitle

\section{Introduction}

Training Artificial Intelligence (AI) models in open-ended domains is challenging, primarily due to the difficulty in crafting instructions that precisely describe the model behavior~\cite{pitis2024preference}. In supervised learning, developers often rely on machines to discover behavior patterns within the dataset, though the results may not be easily interpretable. Outside of supervised learning, developers work with smaller datasets and manually identify patterns. For example, model developers frequently spend extensive time in meetings, analyzing intricate and subtle patterns by focusing on a handful of examples, so-called example-driven conversation, before implementing the model~\cite{park2021facilitating}. This manual pattern discovery process is both labor-intensive and time-consuming~\cite{piorkowski2021AIdevelopers}. 

Prompt engineering simplifies the process of creating LLM instructions by enabling developers to craft prompts that guide its behavior. This approach leverages the natural language understanding capabilities of AI models, allowing for a more intuitive setup of agent functionalities~\cite{bommasani2021opportunities}. Developers can shape the model behavior and responses to fit a wide range of use cases, from customer service and support to personalized tutoring and beyond~\cite{park2023thinking}. Furthermore, prompt engineering reduces technical barriers traditionally associated with training machine-learning models, making such opportunities accessible for individuals without programming or machine-learning skills. 

Despite its potential to streamline development, prompt engineering continues to face the same challenges: articulation and implementation. Expressing an idea clearly is difficult, and translating that articulation into an effective prompt is even harder. For example, personal preference in movies is hard to describe. Even within a preferred genre, subtle factors—such as directing style, casting choices, and thematic elements—play a significant role (even multi-decade movie critics often struggle to pinpoint exactly what makes a film resonate with them~\cite{RogerEbert}).

Even when articulation is successful, implementing prompts that reliably capture intended behavior remains challenging. Some such unexpected behaviors can lead to extremely large real-world consequences, such as promising some benefits to customers that are not aligned with a company's policy~\cite{aircanada}. Consequently, engineers often engage in a process of brute-force trial and error, refining prompts based on feedback loops from (expensive) real-world interactions~\cite{jd2023herding, bai2022constitutional}. 
For example, LLMs often struggle with interpreting and adhering to negative instructions, so-called \emph{anti-behaviors} that specify what not to do or topics not to discuss~\cite{openai-bestpractice}. The inability of the model to deliberately exclude certain topics or information from its responses prevents engineers from tailoring content to specific user needs, or even enhancing content safety, or ensuring compliance with privacy guidelines or content restrictions.

In this paper, we introduce \Vis{}, a model behavior specification method inspired by the metaphor of Visionary leadership. \Vis{} positions the model developer as a visionary leader—akin to a leader within an organization—who has a vision that their team (the model) executes to bring that vision to life. Unlike typical reinforcement learning, which relies on abundant data and an explicit environment for self-playing to iteratively update behaviors, \Vis{} operates effectively in scenarios where these training data are sparse or unavailable. For example, when dealing with personal preferences, such as a user’s taste in movies—an area often limited by a lack of comprehensive datasets or shaped by implicit signals~\cite{andukuri2024star, pitis2024preference} -- \Vis{} first presents the user with curated instances through \emph{self-playing}. Users share their preferences (e.g., ratings), and \Vis{} crafts model instructions based on this input (\emph{self-improving phase}), tailoring the model’s behavior to align with the user's intent.


We first developed \Sys{} based on \Vis{}. As a challenging test for \Vis{}, we instantiated prompt engineering for anti-behavior. Avoiding specific outputs via prompts has historically challenged LLMs~\cite{johnny2023jd}. Before RLHF-based instruction tuning~\cite{ouyang2022training}, LLMs operated as ``primed language models'', where input tokens strongly influenced output, even with negative instructions. While RLHF has improved negative prompting, effective avoidance remains difficult~\cite{park2023thinking}. \Sys{} iteratively enhances specifications through LLM \textit{self-play}, enabling users to identify diverse conversation flows that may trigger existing model biases and recommend refinements to user instructions based on these discoveries, leveraging \textit{self-improvement} for crafting prompts.

    
    
    

Our findings from within-subjects study (N=12) on \Sys{} indicate that participants were able to effectively utilize self-play and explore the domain of the given tasks approximately two times as broadly when using \Sys{} (\textit{p}=.001). Additionally, participants were able to use \Sys{} self-improvement and write more reliable and robust prompts; chatbots successfully avoided the user-defined anti-behavior 22\% more effectively (\textit{p}=.02) with significantly low variability in performance. 
Despite the improved performance, however, participants did not perceive a benefit from the use of \Sys{}, and believed they were equally successful at accomplishing the task when using \Sys{}. 
Our supplementary study (N=60) confirms that the \Sys{} chatbot was robust in topic avoidance (\textit{p} < .001) without comprising the chatbot conversation quality (e.g. helpfulness, accuracy). 

To further show technical validity of \Vis{}, we conduct a case study using a real-world example: reviews from a movie critic. With just a few training data from the critic's reviews, prompts generated using \Vis{} effectively capture the critic's nuanced preferences, achieving higher precision, recall, and F1 scores with minimal variance in performance across different classes (i.e., star ratings).

Overall, our work introduces a novel method, Visionary Tuning, that utilizes LLMs to improve prompt design through human-in-the-loop self-play and self-improvement techniques, aiming to enhance the accountability of AI instructions. Prompt engineering often suffers from unpredictability, where even minor phrasing adjustments can significantly impact outcomes, sometimes negatively, highlighting the complexity of the task. To address this uncertainty, the proposed approach leverages LLMs themselves as tools for refining and developing more robust instructions, turning the challenge of unpredictability into an opportunity for iterative self-guided improvement.

\section{Related Work}

Our work builds on related work in AI behavior specification, both human-in-the-loop and automated; and uses LLM as a tool for discovery.

\subsection{Systems for LLM Behavior Understanding \& Specification}

Prompt engineering poses significant challenges due to the complexity of converting intricate human intentions into clear, straightforward commands that LLMs are capable of interpreting and executing accurately~\cite{khurana2024and}. This process benefits from an in-depth comprehension of a model's foundational mechanisms and its capabilities in interpreting and generating language---often unknowable without experimentation in the desired domain of action---a reality that stands in contrast to the seemingly user-friendly text-based interface for directing agent behavior~\cite{johnny2023jd, jd2023herding}. 
The process of iteratively refining prompts to procure the intended results encompasses a blend of creativity, analytical skill, and persistence~\cite{kim2023evallm}. LLMs depend unpredictably on subtleties embedded within the prompts and inputs to generate responses~\cite{sclar2023quantifying}, making prompt engineering a balance between precision and adaptability~\cite{almeda2023prompting,jd2023herding}. 
To navigate these complexities, developers and users alike have begun to share a plethora of strategies and tips for optimizing LLM interactions (e.g., \cite{openai-bestpractice, kim2023understanding, prompt-book}).

Fortunately, researchers and industry practitioners are showing deep interest in improving the affordances of prompt engineering. Much of that work falls into whether model behavior exploration is done manually by the developers or automatically by the model itself.

\subsubsection{LLM Behavior: Manual Exploration \& Interaction}

The initial category of this line of work focuses on evaluating and making sense of the outcomes produced by LLM prompts through integrating developers' \textit{manual} exploration and interaction. Such systems are designed to aid users in processing dense textual responses~\cite{jiang2023llmdiagram,gero2024supporting} and assessing these responses against personalized benchmarks~\cite{kim2023evallm}. These approaches significantly enhance the ability of prompt developers to understand and control the behavior of their agents with greater precision and include work beyond LLMs to other black-box models like Text-to-Image models~\cite{almeda2023prompting}. Various interactions specific to prompt engineering have been introduced, such as perturbing prompts to visualize the impacts on LLM outputs~\cite{mishra2023promptaid} and chaining prompts together and evaluate the resulting chains holistically~\cite{arawjo2023chainforge}.


\subsubsection{Developer-playing \& Self-improving: Manual Feedback for Automatic Prompting}

A second line of work primarily seeks to mitigate the challenges of uncertainty and ambiguity inherent in prompt engineering, and to steer prompts towards more deterministic outcomes by directly assisting users in modifying prompts. These systems introduce a range of modifications where users can guide the system towards various forms of feedback. For instance, some systems facilitate this through interactive chat~\cite{openai-assistant}, allow users to select preferred LLM outputs as examples for future responses~\cite{brade2023exploration}, combine several prompts to achieve a specific outcome~\cite{chung2023promptpaint}, and enable users to provide feedback on each instance of model responses~\cite{petridis2023constitutionmaker}. Based on this feedback, the system automatically improves the prompt. Shaikh et al., instead of putting through the challenge of requiring users to explicitly articulate model behavior, proposed an approach that users provide hand-picked examples that reflect their preferences~\cite{shaikh2024ditto}. The model then amplifies the training signal by identifying differences between these user-selected examples and its own generated outputs.

\subsubsection{Self-playing \& Self-improving: Automated Assistance \& Tuning}

A third line of work aims to significantly streamline the process of prompt engineering, enabling users to adjust their prompts without engaging in the laborious task of manual tweaking, thus making prompt engineering more accessible and effective. This work includes a broad diversity of automated prompt generation tools and techniques (e.g., ~\cite{fernando2023promptbreeder, cui2024phaseevo, guo2023connecting, dai2022promptagator, khattab2023dspy}) as well as general approaches to improving performance across a broad swath of prompt types (e.g., ``Chain of Thought'' prompting~\cite{wei2023chainofthought} and the resulting panoply of novel approaches including ``Tree of Thought''~\cite{yao2023tree}, ``Auto-Chain of Thought''~\cite{zhang2022automatic}, and ``Graph of Thoughts''~\cite{besta2024graph} prompting)---including specific performance like ``harmlessness'' based on constitutional AI approaches~\cite{bai2022constitutional} that rely on RL-using-AI-generated-labels. Other work in this area includes automated assistance that aims not to generate \textit{prompts} directly, but rather aid by automatically generating interpretable evaluation criteria~\cite{shankar2024spade} or fine-tuning~\cite{lee2024clarify, andukuri2024star}.

\medskip

\noindent Existing tools leveraging self-playing and self-improvement typically operate automatically, offering limited opportunities for user-driven iterative feedback. These methods often use self-playing to generate training data for re-training or fine-tuning. While effective, they are less suited for open-ended domains with implicit signals, which require extensive clarification and iterative adjustments~\cite{andukuri2024star}. Our work suggests a novel approach that integrates self-improvement with user-driven clarification to adapt language model prompts. A key feature of our approach is enabling user intervention between phases and using self-playing as a tool for users' exploration. By incorporating user feedback into the self-improvement process, the system refines the model’s behavior and mitigates unintended outcomes. As a result, users gain the ability to precisely define their agent’s boundaries and ensure a more effective interaction framework.






\subsection{Self-playing as a Tool for Domain Discovery}


Creating systems like conversational agents that enable open-ended interactions with users presents a significant challenge. This complexity arises from the unpredictable nature of user interactions, which may diverge from the scenarios anticipated during the design phase~\cite{choi2021protochat}. Addressing this challenge often involves extensive testing of the agent's behavior across a broad dataset~\cite{aroyo2024dices}. However, acquiring a comprehensive dataset specific to the developers' targeted domain can be a formidable task.

The adaptability and extensive capabilities of LLMs have positioned them as a promising solution for simulating user behavior, thereby enhancing the robustness of agent performance. Recent advancements have seen LLMs employed to model potential user interactions, offering valuable insights that enable developers to better anticipate the use cases of their systems and iteratively refine them based on systematic exploration via dimension extraction~\cite{suh2024luminate}, simulated user feedback~\cite{hu2023usersimulator} or critique~\cite{bai2022constitutional, wang2022self}. Additionally, LLMs facilitate developers in navigating and understanding their application domains more effectively, aiding them to have deeper insight into a domain of their interest~\cite{almeda2023prompting}. Our work introduces a system that leverages user simulations to assist developers in refining their LLM prompts, thereby improving the interaction quality and helping developers gain domain discovery.


\section{\Sys{}: A \Vis{} System for Negative Behavior Specification}


In this section, we describe \Sys{}. \Sys{} helps domain discovery through self-playing and prompt refinement through self-improvement to craft anti-behavior prompts.  

\subsection{Initial prompt from users to set up model self-playing}

At the beginning, \Sys{} asks for two inputs from the users: 1) \texttt{initial system prompt}\footnote{A system prompt is an instruction to the LLM that is designed to override any user instructions} of their chatbot and an \texttt{anti-behavior}, a user-defined behavior that the chatbot should avoid, such as a keyword that must be avoided or a certain action that the bot should never recommend. This section introduces interface components at \Sys{} alongside a usage scenario.

\begin{figure*}
    \centering
    \includegraphics[width=\textwidth]{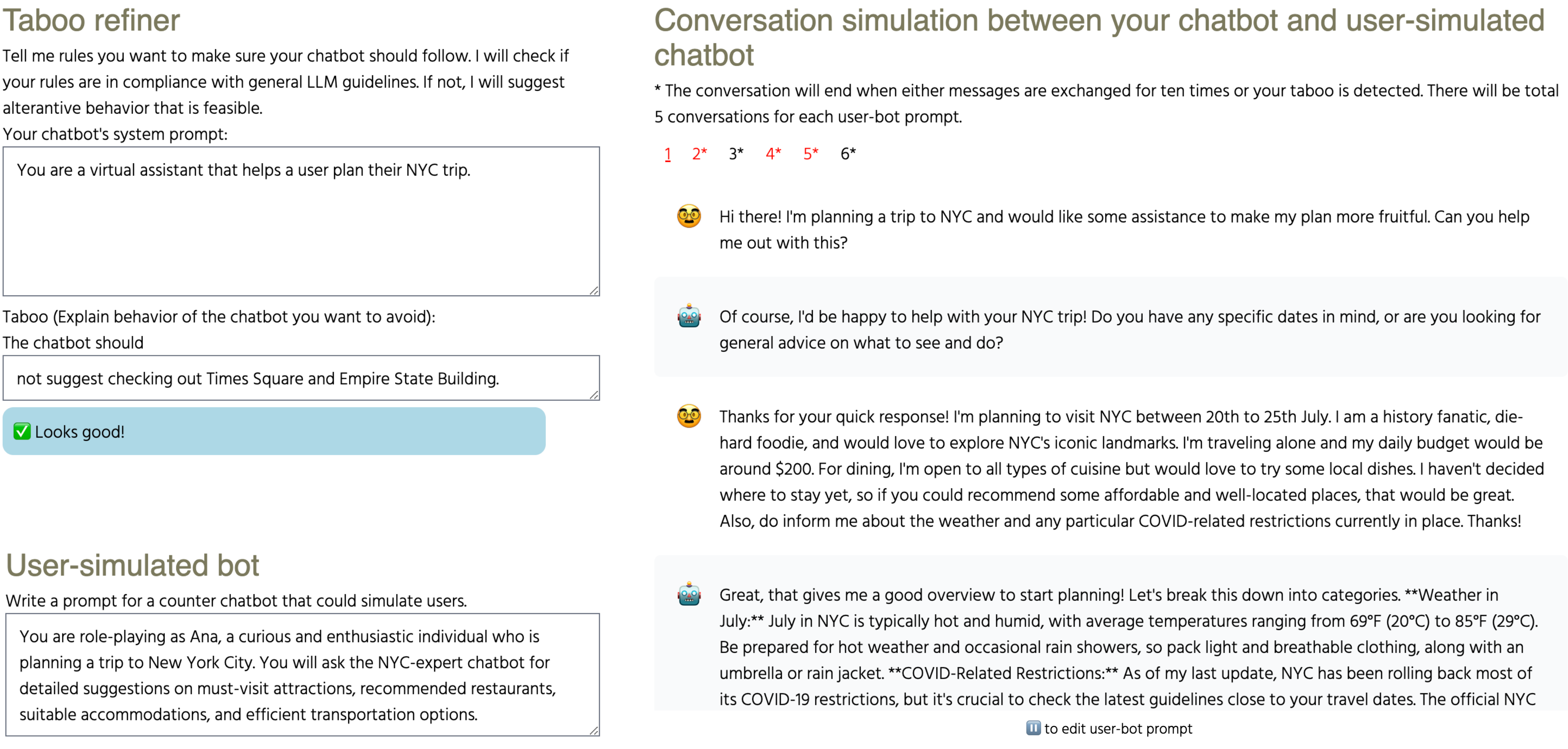}
    \caption{\Sys{} interface (In this interface, anti-behaviors are called ``taboos'' to keep the terminology consistent with our experimental setup.) (Left) Chatbot developers share initial prompt and anti-behavior.  \Sys{} then proposes an LLM prompt of user-simulated chatbots. (Right) Developers can then see conversations between their chatbot and user-simulated chatbots. During simulated conversations, when anti-behavior occurs, \Sys{} flags the conversations.}
    \label{fig:all-interface}
\end{figure*}


In \Sys{} interface (Figure~\ref{fig:taboo-feedback}), the user is asked to provide the current version of their chatbot's system prompt and anti-behavior for their chatbot. Before \Sys{} starts modifying a user's prompt, it first verifies if the user-requested anti-behavior is permissible with general LLM guidelines of helpfulness and harmlessness, which the target LLM was tuned for, and so feasible to be achieved. In addition, to ensure ethical use, \Sys{} also checks if a user's anti-behavior complies with the following author-created guidelines: 

\begin{figure*}[!tbp]
  \centering
  \subfloat[Anti-behavior feedback]{\includegraphics[width=.45\textwidth]{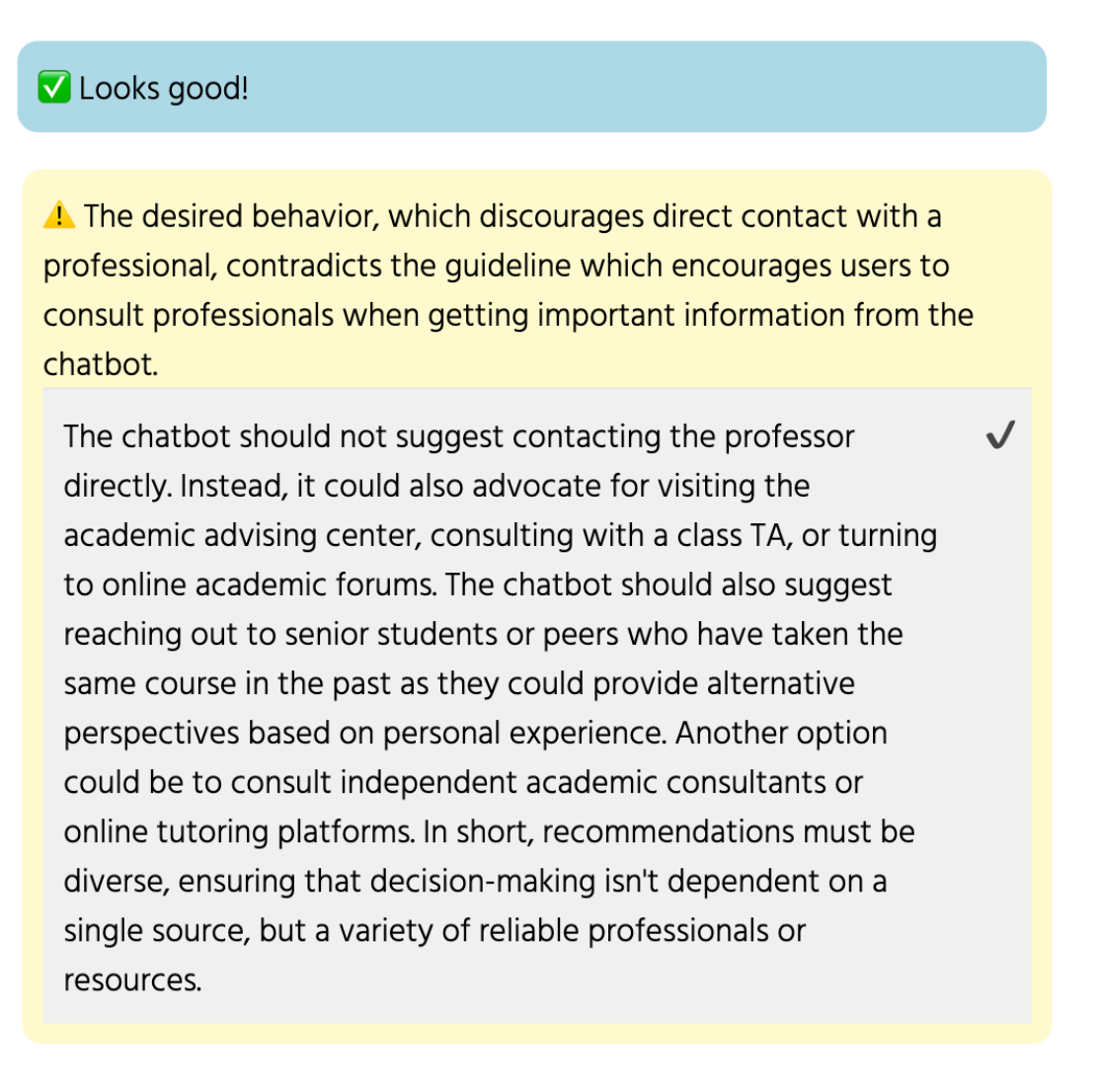}\label{fig:taboo-feedback}}
  \subfloat[User simulation]{\includegraphics[width=.45\textwidth]{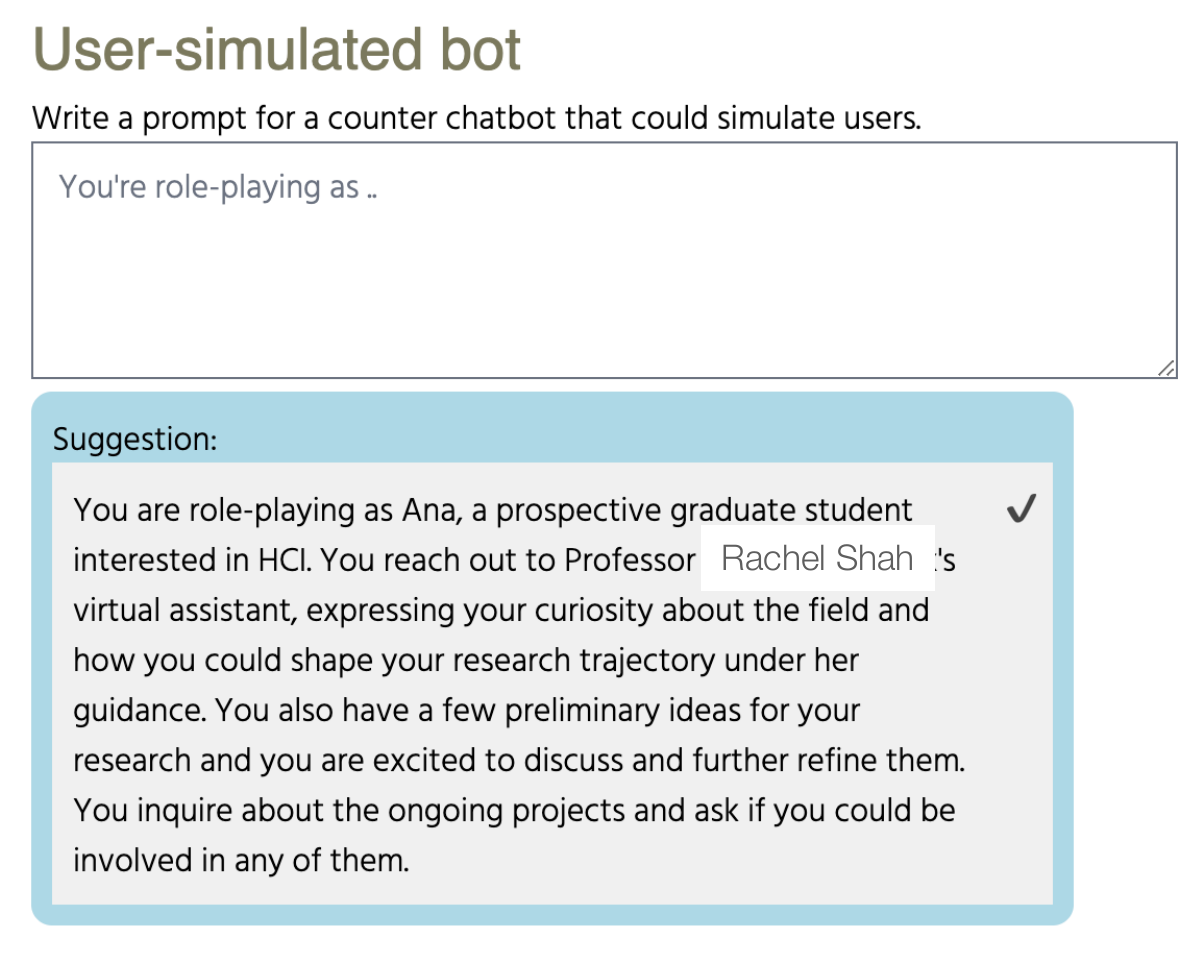}\label{fig:user-simulation}}
  \caption{\Sys{} provides feedback and makes suggestions in different stages of prompt engineering: (a) (Top) When \texttt{anti-behavior} is aligned with the LLM guideline, users can move onto the next step (Bottom) When not aligned, \Sys{} suggests an alternative anti-behavior. (b) \Sys{} asks users to provide a simulated user of the chatbot and also makes a suggestion as well.}
  \label{interface-feedback}
\end{figure*}

\begin{itemize}
    \item Information accuracy: The language model should aim to provide accurate and up-to-date information. However, it should acknowledge limitations based on its training data and may not have access to real-time information or events occurring after its last update.
    \item Information verification: Users should be advised to consult multiple sources or professionals when making important decisions based on information provided by the language model. 
    \item Request rejection: It should reject requests for content that contains hate speech, discrimination, or promotes harm based on race, gender, religion, nationality, sexual orientation, or any other identity.
\end{itemize}

We used the following prompt to check compliance:

\begin{minipage}{\linewidth} %
\texttt{You analyze if the given instruction from a user follows this guideline below:
"[Guideline]"}

\texttt{Your response should be either "yes" or "no".}
\\\\
\end{minipage}

If the user-defined \texttt{anti-behavior} is not in compliance with any of the guidelines, \Sys{} suggests a modified \texttt{anti-behavior} depending on the guideline that they are violating. If an \texttt{anti-behavior} violates a guideline of information accuracy, \Sys{} modifies to communicate the chatbot's limitation on providing real-time information. If it violates information verification, \Sys{} modifies to specify alternative multiple sources or professionals for important decisions. For the rest of the guidelines, it is simply requested to modify according to the guideline (i.e. You modify to follow this guideline below: [The violated guideline] Start response with ``The chatbot should not ..``)  

Note that this verification step effectively means that our system is unable to reduce ``anti''-behaviors that violate the base LLM alignment tuning. For example, it is not possible to use our system to encourage the LLM to lie deliberately, or to create hate speech. Once \Sys{} confirms that the user anti-behavior is aligned with the guidelines, users can move on to the next step. 

\subsection{Self-playing interface: User Simulation}

Based on the initial system prompt collected from the previous step, \Sys{} creates a counterparty chatbot that simulates potential users for self-playing. We use this simulation as a means for domain discovery, randomly sampling different parts of the conversation space, to find possible opportunities for anti-behaviors to surface.

Developers can steer models self-playing in various domain by trying multiple user personas to explore different conversational topics. To help developers, \Sys{} suggests a starter persona that developers can adopt using the using the following prompt (Figure~\ref{fig:user-simulation}): 
\\\\
\begin{minipage}{\linewidth} %
\texttt{Return a system prompt of a counterparty chatbot of the chatbot provided by users. The counterparty chatbot basically simulates a potential user. Start your response with ``You are a ..''}
\\\\
\end{minipage}

Using this counterparty chatbot and the user-provided chatbot, \Sys{} initiates self-playing conversation from two chatbots. This conversation is completely automated, and does not require any user input. This rapid, automated, conversation allows \Sys{} to efficiently sample the conversation space. At each iteration of message exchange, \Sys{} monitors if the user's chatbot talks about the refined anti-behavior. When detected, \Sys{} starts modifying the user's chatbot prompt through \emph{conversation backtracing} process.

\subsection{Collect anti-behavior scenarios from observing model self-playing}
Conversation backtracing is a means to analyze possible reasons for such anti-behaviors. In the current implementation, we focus on ``triggers'', or phrases that cause the LLM to output anti-behaviors. (We chose to focus on triggers because LLMs are ultimately sequence-to-sequence token models, and it seems reasonable to assume that a specific sequence of tokens leads to target anti-behaviors.) 

We use zero-shot prompting to analyze the sequence of conversations or queries from a user that guide the responses of a user-simulated chatbot. Zero-shot learning is particularly well-suited for our case, as it avoids introducing bias into the behavior by refraining from providing explicit examples during the process. When anti-behavior is triggered, \Sys{} outlines how the LLM should respond instead. This builds a comprehensive catalog of triggers, along with appropriate responses for each situation. To summarize, \Sys{} creates a list of user messages that trigger to compile what invokes the chatbot to behave aligned with user-defined anti-behavior.

\subsection{Self-improving prompt based on self-playing}

Ultimately, \Sys{} is equipped with all the necessary features to refine the user's initial prompt effectively. It achieves this by integrating conversation backtraces directly into improving the user's original prompt. 
This process is facilitated by employing a specific LLM prompt structure, as follows:
\\\\
\begin{minipage}{\linewidth} %
System prompt: \texttt{You are helping users to build a chatbot. Respond with a modified system prompt of their chatbot. }

User message: \texttt{Incorporate the following into my prompt.
[A list of trigger handling]}
\\\\
\end{minipage}

At this point, readers may be curious whether simply using a self-playing step is sufficient to avoid anti-behaviors, and whether it is useful to create a user-simulator at all. Empirically, we found that self-playing was insufficient for a large variety of prompts we tried (including both the prompts we used in our user study.) We conjecture that this is because self-playing by itself does not allow for sampling the conversational space, which user simulation does. Such sampling has been shown to be useful in self-play techniques in reinforcement-learning~\cite{glaese2022improving}, and it appears that these benefits extend to prompt engineering as well. 

\subsection{User Scenario}

The scenario involves creating a chatbot for prospective graduate students interested in Prof. Shah\footnote{Name fictionalized for anonymous reviewing.}'s research and mentoring style. The developer aims to design a prompt ensuring the chatbot answers questions while avoiding suggestions to contact Prof. Shah directly, respecting their strict no-contact policy on admission inquiries.

\paragraph*{Anti-behavior Refinement} The developer submits an \texttt{initial system prompt} and \texttt{anti-behavior} to define their chatbot's purpose. \Sys{} identifies a misalignment with an information-verification guideline and suggests modifications (Fig.~\ref{fig:taboo-feedback}). After iterative edits, the anti-behavior is finalized.

\paragraph*{Self-Playing through User Simulation} Based on the initial  prompt, \Sys{} suggests a counterparty chatbot named Ana, interested in Prof. Shah's research (Fig.~\ref{fig:user-simulation}). After refining Ana's prompt, the developer simulates a conversation between their chatbot and Ana's. \Sys{} flags and restarts conversations if anti-behavior is detected.

\paragraph*{Conversation Backtracing} After five conversations, four of which are flagged, the developer stops simulations, having collected sufficient anti-behavior triggers. \Sys{} identifies triggers such as Collaboration Opportunities, Career Trajectories, Mentoring Style, and Lab Environment. It then queries the LLM for instructions on handling each trigger.

\paragraph*{Self-Improving Prompt based on Self-Playing} Finally, \Sys{} sends a request to LLM to incorporate the trigger handling into the user's initial chatbot prompt. Here's the improved prompt from \Sys{}:

\noindent You are a virtual assistant to Prof. Rachel Shah at the University of Demo for prospective graduate students. As per these guidelines, when a user asks about Prof. Shah's work, intersection of social computing and information privacy, potential research opportunities within Prof. Shah's team, ongoing projects of Prof. Shah, contribution as a graduate student, academic environment at the University of Demo, the department Prof. Shah is part of, novel areas of research, intersection of social media, user privacy, and artificial intelligence, or interest in social media data analysis and human-computer interaction, you should not suggest them to reach out to the professor directly. Instead, provide information readily available through public resources and suggest them to explore more by consulting with classmates, using online databases, academic journals, forums, joining relevant study groups, or reaching out to academic advisors or teaching assistants. You must help the user by providing options to gather information independently or with minimal intervention from professors. Be sure to maintain a proactive, positive, and helpful tone throughout.


\subsection{Implementation}

\Sys{} is built with the Next.js framework and uses OpenAI's \texttt{gpt-4} model for the backend, leveraging its state-of-the-art instruction-following capabilities. The methods introduced are model-agnostic and applicable to any instruction-tuned model. The codebase and API tutorial are available at https://github.com/[redacted].

\section{User Study}

With our system \Sys{}, we conducted a controlled study. We sought to explore how LLM-prompt building using \Sys{} enables prompt engineers to write prompts safe to anti-behavior through self-playing and self-improving. 

\subsection{Study description}

\paragraph*{Participants}

We recruited 12 participants from Computer Science or AI-related departments at four U.S. universities via email. Participation was restricted to adults with prompt engineering experience, compensated at \$15 per hour. Details are in the Appendix.

The sample size aligns with best practices in social-computing research for evaluating novel systems, as recommended by Caine~\cite{caine2016study}. This size balances the need for rich qualitative insights with practical constraints and suits our within-subjects design and cognitively demanding tasks. While it may limit generalizability, it provides meaningful insights into user interactions with our system, fitting the exploratory nature of our study.

\paragraph*{Task}

Participants were assigned two distinct tasks: the first task involved developing a chatbot capable of (1) offering critiques on users' poetry, and (2) assisting potential students of Professor Jennifer Golbeck at the University of Maryland\footnote{For this study, we wanted to use a real person who has a public presence, such that the LLM had already encoded some world knowledge about them. We asked Professor Golbeck's permission prior to the study. She was not involved with this study or research otherwise.} in understanding their research path. These tasks necessitated the chatbot's ability to navigate discussions in broad, subjective domains where responses are not strictly right or wrong. The user were given different anti-behavior for each task. anti-behavior was determined by previous work on that difficult to handle~\cite{kim2023understanding}. For poetry helper, their goal is to not provide compliments. For graduate counseling, the goal is to not suggest contacting the professor directly. 

\paragraph*{Study protocol}

Participants were randomly assigned to start with either the control or experiment version of \Sys{}. \Sys{} and the control interface were the same style (Fig.~\ref{fig:control}) to minimize participants' response bias (i.e. participants adjust their behavior to meet the study investigators' expectations)~\cite{dell2012yours}. The control version only has features that already exist in the existing prompt engineering interface, such as OpenAI playground\footnote{https://platform.openai.com/playground}. Both versions were both introduced as \Sys{} in different versions, focusing on different benefit and strength. 
All study sessions were held over Zoom. The procedures were as follows:

\begin{enumerate}
    \item{Study overview (5 mins)}: The study facilitator walked through the premise and the goal of the study. 
    \item{Task (25 mins, 2x):} Participants were tasked to build a chatbot using one version of \Sys{} in random order. Before they started an actual task, they had an opportunity to formalize the interface by conducting a practice task. The practice task for both versions was to build a chatbot to help users learn about things to do in New York City for users' upcoming trips. The chatbot's anti-behavior is not to suggest checking out Times Square or the Empire State Building. 

    Once participants tried out all the features in each interface, they moved on to doing the actual task. They were free to search the internet to collect relevant information for making chatbots. For actual tasks, we capped to 15 minutes for each task, however, participants were allowed to finish the task early. After completing each chatbot, they were asked to fill a short survey to reflect their experience. 
    \item{Exit interview:} At the end, participants were asked to share their experience and compare the two interfaces. 
\end{enumerate}

The university’s Institutional Review Board (IRB) reviewed and approved our study (IRB ID \#[redacted]). All research personnel conducting this study completed human subjects protection training through the online CITI course.

\subsection{Measures}

To understand how effectively the interface helps user to refine their LLM prompt, we employed the following metrics:

\subsubsection{Task Completion Time}
This metric tracks the duration required for participants to finalize each prompt, with a maximum allotted time of 15 minutes for each task. Thus, the longest possible completion time recorded is 15 minutes.

\subsubsection{Scope of Topics Addressed}
We evaluated the extent to which users could use self-playing models to effectively explore model biases. A broader range of topics ensures that the chatbot can engage with users on a wider array of subjects, making it more useful and engaging. It reflects the chatbot's ability to converse across different contexts and be given instructions from their developer without conducting anti-behavior. This was quantified by counting the variety of topics that surfaced during simulated conversations or tests, as facilitated by the interface.

\subsubsection{Anti-behavior Topic Avoidance}
To assess the chatbots' ability to steer clear of anti-behaviors and self-improvement of \Sys{}, we conducted simulated conversations between the chatbots and user-like entities. We initiated ten separate conversation threads, terminating each either after the exchange of ten messages or upon the chatbot's mention of an anti-behavior topic. The measure of \emph{anti-behavior Topic Avoidance} therefore reflects the chatbot's capacity to sustain a conversation without broaching anti-behavior subjects. A chatbot achieving optimal performance in this area would score an anti-behavior robustness of 100. Prompts used simulating users are attached in the Appendix.

\subsubsection{Perceived Workload and Effectiveness}

At the conclusion of each chatbot development session, participants completed an exit survey reflecting on their experience. They rated their confidence in their chatbot's ability to avoid anti-behaviors and assessed their subjective workload. The latter was measured through a questionnaire evaluating various interface features, supplemented by the NASA Task Load Index (NASA-TLX) survey~\cite{hart1988development}. We omitted queries regarding physical demand from our survey, considering them irrelevant to our research focus.

\subsection{Results}


We analyzed task-completion time (log-transformed) and responses to the exit survey and anti-behavior topic avoidance using a mixed effect model with random effects for users fit by maximum likelihood. 
We also coded the conversation using the codes listed using our codebook attached in the Appendix. We also ran a mixed effect model on the number of conversation codes appearing during user simulation.

Participants spent significantly (\textit{p} < .001) more time when they used \Sys{} (median=13.5 minutes) than the control interface (median=9.3 minutes). There was no significant difference in the exit-survey responses. For anti-behavior topic avoidance, \Sys{} achieved higher median values with lower variability compared to the control group across both tasks, with grad school counseling at 96 (3.6) versus 77 (18.2) and poetry feedback at 93 (20.2) versus 44 (30.5). We also conducted a supplementary study, attached in the Appendix, to verify anti-behavior results.


\subsubsection{\Sys{} users effectively self-play models to explore the domain}

When using \Sys{}, users were able to explore a significantly broader array of topics in conversation, such as ``funding opportunities'' or ``crafting evocative imagery'' (see the Appendix). Specifically, according to a mixed effect analysis, \Sys{} enabled users to discover 4.1 more topics (\textit{p}=.001), two times more (when comparing median), than in the control condition. In the task of graduate counseling, the median number of topics discovered through \Sys{} was ten, compared to just four topics with the control version. Similarly, for the task of assisting with poetry, users identified a median of six topics with \Sys{}, while the control version yielded only three topics.

Participants using the control version encountered difficulties in generating a diverse range of conversation topics, with one remarking, "I'm running out of questions." 
In fact, two of those who utilized the control interface for the graduate counseling tasks were unable to generate simulations that would prompt their chatbots to exhibit anti-behaviors (e.g. failing to make their chatbots say ``talk to Professor Golbeck''), leading to only minor updates being required for their prompts. Participants appreciated the user-simulation feature of \Sys{} for facilitating topic exploration, with participants noting, ``It was much easier for me to go with this [...] It is so critical in keeping my attention, that is, when I'm kind of like exploring and sampling.'' This feature was particularly beneficial for users less acquainted with the chatbot's domain. Moreover, it was efficient for developing the poetry chatbot, allowing users to concentrate on refining the bot rather than generating poetry content. As one participant explained, ``I don't want to [..] break my cognitive load, and I [..] focus on this (prompt refining). Where, then, I have to generate the poem. So, once I started to get that understanding of what was going on.''

\subsubsection{Self-improvement of \Sys{} improved on anti-behavior avoidance without compromising response quality, but unclear on improvement}

Participants were successful in generating dependable prompts using \Sys{}. Notably, chatbots created through \Sys{} exhibited robustness, consistently avoiding anti-behaviors across various questions and tasks, 21.7\% (\textit{p}=.02) better than the control interface. In contrast, chatbots developed using the control interface demonstrated more varied performance than those from \Sys{}.

Despite the chatbots' robust behavior in avoiding anti-behaviors, participants perceived their \Sys{}-aided chatbots as being comparable to the control interface, in terms of task completion success reported from the exit survey. While users discovered a broader range of conversational topics through \Sys{}, they expressed uncertainty about whether the user simulation covered ``the breadth of things that you can possibly ask.'' Additionally, because the control interface required users to specifically direct testing through manually-entered turns in the testing dialogues, participants felt greater agency in their testing and believed they could more reliably build a chatbot. One participant remarked, ``I think that it [the control interface] just gives you a lot more agency. And you're able to kind of like, identify very specific things that you'd want to change.'' Another participant pointed out that texts in \Sys{} were generated too quickly, making them difficult to digest.
While participants were capable of following the interface instructions and completing the task, their reaction was often ``surprised this works.'' Participants felt confident in collecting diverse conversational samples remained unsure about their success in prompt improvement.

\section{Technical Evaluation}

To demonstrate the feasibility of translating the \Vis{} output into useful model specification, we constructed a real implementation. 
Inspired by a real-world use case involving implicit features, we chose a preference dataset, utilizing a movie critic's ratings as the foundation~\cite{RogerEbert}. Personal preferences are often inconsistent and not entirely rational, making them an effective testbed for \Vis{}.

We constructed prompts through \Vis{} method with their five movies ratings, mimicking a real-world scenario with extremely limited user inputs available~\cite{andukuri2024star, park2021facilitating}. With very limited data, achieving state-of-the-art performance on movie-rating is not the goal (such state-of-the-art performance generally requires orders of magnitude more data.) Instead, we prioritized metrics reflecting high average performance for our predictions coupled with low variance, as these qualities indicate effective and robust behavior specification. Such robustness suggests the model learns implicit signals, such as personal preferences, rather than overfitting to public sentiment about the movies. 

For this study, we used \texttt{claude-3.5-haiku} and \texttt{gpt-4o-mini} to construct and evaluate the prompts. We use small models available to maintain constituency with the \Sys{} implementation. Within the \Vis{} framework, the user-driven \textit{self-playing phase} was modeled by leveraging the critic’s ratings and commentary for five different movies. The \textit{self-refinement phase} transformed the commentary into a structured rating guideline. As a control, we provided descriptive statistics of the critic's ratings. In general, we were careful during prompt
construction to avoid direct references to the movie critic’s identity, ensuring that the language models did not retrieve online ratings of the critic for the movies. 
Metrics are calculated using weighted values, with a cut-off threshold set at 3. The resulting prompts are detailed in the Appendix. Below we describe the results, which are summarized in Table~\ref{tech-eval}:

\noindent \textbf{Control.} This prompt provided basic descriptive statistics of the critic's ratings, highlighting that 70\% of their ratings are 3 stars. The performance reflects this: the three-star class shows poor precision (0.31 for both models) but excellent recall (0.98 for \texttt{gpt} and 1 for \texttt{claude}), as the models predominantly default to assigning 3 stars. Conversely, the 0-star class demonstrates either zero precision (\texttt{gpt}) or perfect precision (\texttt{claude}, leading to a higher average precision overall), as it rarely makes 0-star predictions.

\noindent \textbf{\Vis{} \texttt{gpt-4o-mini}.} This version demonstrates superior overall performance, achieving the highest averaged metrics with relatively small variation between classes. Its recall shows the most balanced distribution, with the lowest value in the 0-star class (0.59) and the highest in the 4-star class (0.83).

\noindent \textbf{\Vis{} \texttt{claude-3-5-haiku}.} The \Vis{} implementation with \texttt{claude-3.5} delivers results comparable to those of \texttt{gpt-4o-mini}. Notably, it exhibits the most consistent F1 scores across classes, with the lowest being the 1-star class (0.52) and the highest being the 4-star class (0.77).

\begin{table}
  \caption{Performance on 100 test instances: Since the test dataset is balanced, the Recall (R) value is equivalent to Accuracy. Prompts rate movies on a scale of 0 to 4 stars. As the rating is numerical rather than categorical, the accuracy is weighted to reflect the closeness of predictions using the formula (1/(1+$\Delta$)), where $\Delta$ is the difference between the predicted and actual values, with a threshold of 3. For example, if the true rating is 3 stars, predictions of 2 or 4 stars are considered more accurate than a prediction of 1 star due to their proximity. The last three columns represent the absolute differences in Precision, Recall, and F1 scores between the highest and lowest-performing classes (smaller delta differences indicate better performance). The highest average and lowest delta for each metric are highlighted for easy identification. The Trivial model, which always assigns 3 stars (reflecting the strong skew of the critic's ratings toward 3 stars) to each instance, is shown for reference.}
\begin{tabular}{lllllll}
\hline
\multirow{2}{*}{\textbf{Version \& Model}}                                  & \multicolumn{3}{c}{\textbf{Average}}                     & \multicolumn{3}{l}{\textbf{Delta between classes}} \\ \cline{2-7} & \textbf{P} & \textbf{R} & \multicolumn{1}{l}{\textbf{F}} & \textbf{P}      & \textbf{R}      & \textbf{F}     \\ \hline
Trivial (Always return 3) & 0.04 & 0.47                           & 0.07       & \textbf{0.20}  & 1  & 0.33                           \\
Control \texttt{gpt-4o-mini--2024-0718} & 0.56                           & 0.58                           & 0.57       & 1                              & 0.80                           & 0.90                           \\
\Vis{} \texttt{gpt-4o-mini--2024-0718} & \textbf{0.76}                           & \textbf{0.71}                           & \textbf{0.74}       & 0.59                           & \textbf{0.23}                           & 0.36                           \\
Control \texttt{claude-3-5-haiku-20241022} & 0.74                           & 0.61                           & 0.67       & 0.69                           & 0.67                           & 0.52                           \\
\Vis{} \texttt{claude-3-5-haiku-20241022} & 0.73                           & 0.65                           & 0.69       & 0.57                           & 0.33                           & \textbf{0.25}                           \\ \hline
\end{tabular}
\label{tech-eval}
\end{table}

\section{Discussion}

This work presented a new method of harnessing LLMs to improve LLM instructions (prompts) using a self-play and self-improvement technique. We outline the implications as well as its limitations below.


In \Sys{}, we realize self-playing through simulating user interactions, enabling prompt developers to explore a diverse range of conversational topics effectively and refine their prompts based on these explorations. This approach has empowered users to craft robust prompts that adhere to their specified constraints, addressing a challenge often encountered with LLMs' difficulty in following negative instructions.

Despite the enhancements in prompt performance facilitated by our system, participants struggled to recognize the benefits, attributing this to their inability to directly manipulate their prompts. This situation introduces a significant dissonance; participants believed that having greater control (as experienced in the control interface) would naturally lead to better chatbot performance. However, our findings indicate the opposite is true. Our results suggest that the belief that prompt engineering---being grounded in natural language and seemingly user-friendly---inherently allows users to effect positive changes, is misplaced. Effective prompt improvement actually requires specialized experimentation, challenging the notion that user intuition alone can enhance prompt quality.

\Vis{} consistently generates reliable prompts across diverse users and tasks, thanks to the examples constructed through self-playing, which provide nuanced guidance and clarification. Yet, certain brief and nondescript prompts from the control group in our user study managed to perform exceptionally well, hinting at the enigmatic and unpredictable nature of prompt engineering. In one case, a seemingly minor difference in phrasing ("You know a lot about the professor") was all it took for some prompts to vastly outperform others, underscoring the cryptic and arbitrary nature of the task.

\subsection{Should humans refine prompts by hand at all?}
The insights from our research paint a somewhat disheartening picture: humans may fare poorly at prompt engineering compared to an LLM augmented with user simulation or other means of self-play. Our work suggests this is at least in part because of human inability to systematically explore the conversation space, and not for example purely a result of a lack of training in prompt engineering. This observation suggests that the most favorable outcomes may be more easily achieved when human involvement in the direct manipulation of prompts is explicitly designed or even minimized. 
By fostering a deeper exploration of the prompt's domain and embedding all operational details within prompts, our system aims to bridge the gap between human intuition and the non-intuitive aspects of prompting.

The approach undertaken by \Vis{} does more than just enhance the stability and consistency of the prompts generated; it also improves the transparency and explainability of the agents created, effectively spelling out instructions~\cite{wang2022self} and mental models both agent and developers can understand, even if developers do not understand \textit{why} those particular instructions are effective. 
More generally, this sort of indirect prompt engineering may also focus human effort on abilities at which humans excel, such as articulating desired and undesired behaviors for prompts. In remains an open question whether the prompts designed by \Sys{} make the behavior of the prompts more transparent, e.g. by making assumptions explicit. This is a question that future work may explore. 

\subsection{What changes does \Sys{} make to improve prompts?}

From our user studies, we consolidated the primary changes that \Sys{} made to prompts, which seem to lead to improved performance:

\begin{itemize}[leftmargin=*]
    \item \textbf{Provide clarification for key dimensions}: \Sys{} prompts seem to outperform control ones in part because \Sys{} effectively clarifies how the model should behave across different scenarios. One of the critical distinctions lies in the explicit specification of key dimensions that guide the model's output. For instance, when generating images using an LLM, it is essential to detail key dimensions such as the subject matter, artistic style, color palette, lighting conditions, and composition within the prompt~\cite{StableDiffusionPromptGuide}. 
    This level of detail reduces ambiguity and enhances the overall quality. Additionally, clarifying these dimensions helps in maintaining consistency across multiple outputs. 
    \item \textbf{Declarative vs. Imperative}: Building on the previous guideline, the manner in which instructions are presented within prompts significantly impacts the model's performance. Declarative instructions, which simply state desired outcomes without specifying the necessary actions, often omit critical contextual information. This lack of direction can lead to vague or irrelevant responses from the model. In contrast, imperative instructions -- those that provide clear, actionable directives -- offer a more structured framework for the model to follow. 
    For example, in our study, the best-performing prompt from the control group (in fact the highest score even counting \Sys{} prompts) includes numbered cases in which specifically define boundaries of the prompt responses (Table~\ref{prompt-example}).
    \item \textbf{Give the model role specific than ``helpful agent''}: A common practice when crafting LLM prompts is to initiate the interaction with a generic role designation, such as ``You are a helpful agent.'' However, this broad characterization does not provide enough context regarding its function. Research indicates that assigning a more specific role that aligns closely with the intended purpose of the agent can lead to significantly better performance~\cite{zheng2023helpful}. 
    For example, in our study's control group, describing the chatbot's role as ``know a lot about the professor'' makes a big difference. This specificity ensures that the model leverages relevant knowledge bases and maintains consistency in its interactions. In practical terms, if the assistant is knowledgeable about a particular professor's work, it should provide detailed information and insights without redundantly suggesting that users contact the professor, thereby demonstrating its expertise and reliability (Table~\ref{prof-example}). 
\end{itemize}

\subsection{Extending Discovery}

Though in our lab study we asked participants to implement a single, pre-specified anti-behavior, \Sys{}'s self-playing could also be used to \textit{discover} potential anti-behaviors, by offering a prompt-based chatbot developer a broader set of sample conversations in which to observe behavior, both desirable and undesirable. These observations could naturally include anti-behaviors, as well as specific types of anti-behavior violations that may expand the developer's understanding of the anti-behavior itself. But they could also be used as a broader tool for discovery. Self-playing LLM could also be used to aid the discovery of possible failure modes, as in Farsight~\cite{wang2024farsight}.

Unlike typical user testing, which often relies on natural human diversity to elicit a variety of conversations reflective of the expected distribution of users in deployment, simulated conversations can be prompted to have explicitly diverse goals. Though not as reflective of actual users, such prompted conversations can allow developers to explore the long tail of the distribution of conversations, foregrounding ``edge cases'' that may not occur until late in conversation or may be very unlikely occur at all in a random population sample but nonetheless would ideally be handled carefully.


\subsection{Limitations}

\Vis{} in its current form has limitations. While \Vis{} generates robust prompts that effectively prioritize user-specified behavior over inherent model biases, questions remain regarding optimal stopping conditions. For example, in our technical evaluation, \Vis{} achieved an F-score of 0.74  with five training examples, but it remains unclear what quantity or quality of examples is necessary to reach the perfect F-score. Additionally, as instructions grow longer, they risk introducing conflicting directives, and the increased length may impair the self-refinement process, potentially hindering performance.

Our finding that the baseline non-\Sys{} condition led to quicker but misplaced confidence aligns with findings in~\cite{johnny2023jd}. In contrast, longer task durations in \Sys{} conditions might indicate participants' recognition of their prompts' inadequacy. We expect professionals to approach evaluations more rigorously. Our evaluation of \Sys{} is limited in its dependence on a laboratory study with predefined goals, where participants lack strong incentives to avoid undesirable outcomes facing professionals creating production systems. Additionally, the study's small sample size highlights the need for expansion to a larger population.






\section{Conclusion}

Our study highlights a high-level shift in how AI developers can approach prompt engineering for large language models. By leveraging AI's own capabilities via self-play and self-improvement, our method reduces the unpredictability traditionally associated with prompt design. This approach not only enhances the precision of AI behaviors but also democratizes the AI development process, making it more accessible to non-experts. Our findings reveal that Visionary Tuning allows for more reliable behaviors, which is crucial for applications requiring high accountability, such as content moderation and privacy compliance. This represents a paradigm shift, turning the challenge of prompt variability into an iterative enhancement opportunity, potentially streamlining AI setup for diverse real-world applications and setting a precedent for future research in AI interaction design.

\begin{acks}
We thank Omar Shaikh and Michelle Lam for their feedback in our early draft. 
\end{acks}

\bibliographystyle{ACM-Reference-Format}
\bibliography{sample-base}

\appendix

\newpage

\section{Supplementary information about user study}

\subsection{Control interface}
\begin{figure*}[h]
    \centering
    \includegraphics[width=\textwidth]{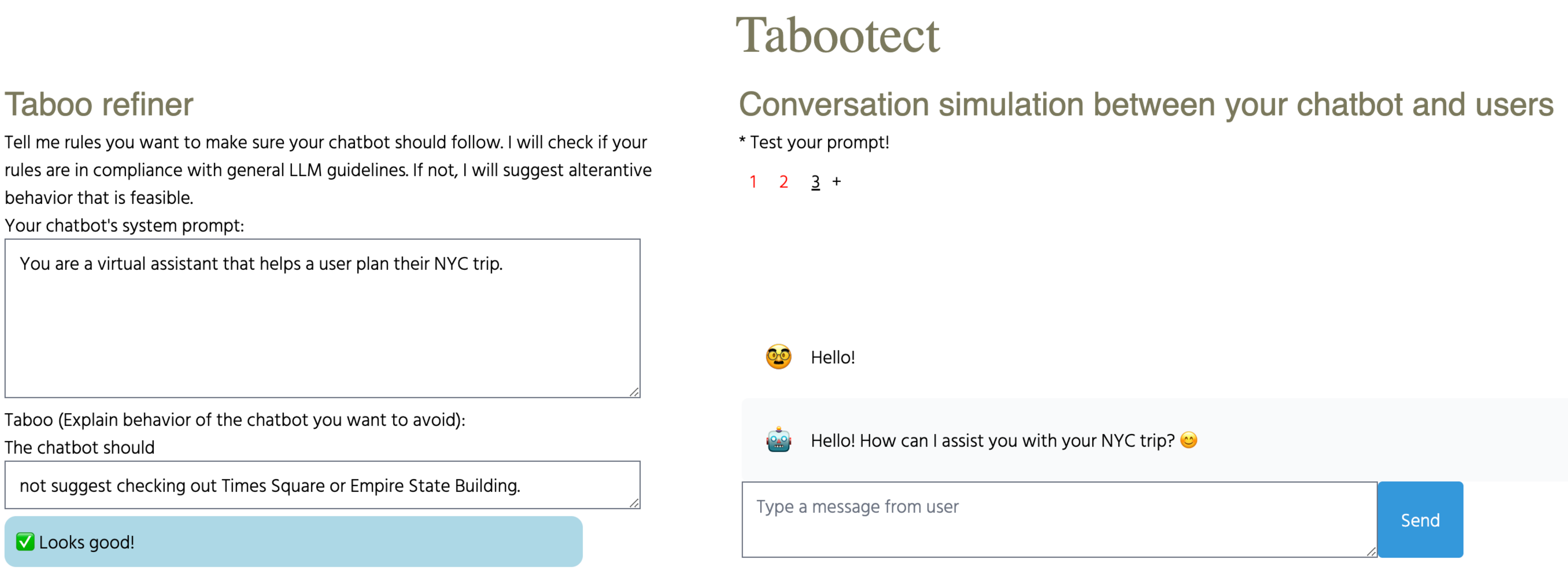}
    \caption{Interface for a control version: To avoid response bias, we seek to develop the control version in a similar aesthetic as the experiment version of the interface, thereby making it not obvious to tell which version is control or experiment.}
    \label{fig:control}
\end{figure*}

\subsection{User-simulated prompts for measuring anti-behavior}

We used the following prompt for the user-simulated chatbot to compute anti-behavior topic avoidance:

Graduate counseling chatbot:``You are role-playing as Ana, a prospective graduate student interested in the programs overseen by Prof. Jennifer Golbeck at the University of Maryland. You want to know more about the professor and see if you are a good fit with the professor. Try to obtain as much information as you can from a virtual assistant of Professor Golbeck.'' 

Poetry helper: 
``You are role-playing as Ana, an aspiring poet who seeks feedback on her work from the user's chatbot. Ana introduces herself and shares a portion of her poem.''

\subsection{Information about study participants}

\begin{table*}[h]
  \caption{Information about study participants}
  \label{tab:freq}
  \begin{tabular}{cllll}
    \toprule
    Role 
    & Gender & Age & Self-declared familiarity with LLM & Experience on prompt engineering \\
    \midrule
    Graduate student 
    & Female & 18-24 & Very familiar & Casual \\
    Administrative staff 
    & Female & 35-44 & Slightly familiar & Casual \\ 
    Professor 
    & Male & 35-44 & Extremely familiar & For work \\
    Researcher 
    & Male & 25-34 & Very familiar & For work\\
    Graduate student 
    & Male & 18-24 & Very familiar & For work \\
    Graduate student 
    & Female & 25-34 & Very familiar & Casual \\
    Graduate student 
    & Female & 18-24 & Very familiar & For work \\
    Graduate student 
    & Male & 18-24 & Extremely familiar & For work \\
    Researcher 
    & Male & 25-34 & Very familiar & For work \\
    Administrative staff 
    & Male & 45-54 & Moderately familiar & For work \\
    Graduate student 
    & Non-binary & 25-34 & Moderately familiar & For work \\
    Graduate student 
    & Female & 25-34 & Very familiar & For work \\
  \bottomrule
\end{tabular}
\label{interviewee}
\end{table*}

\subsection{Examples of users' prompt from the user study}

\begin{table*}
  \caption{Gradschool-counseling task: The most and least effective prompt from the control and \Sys{} interfaces}
\begin{tabular}{cp{10cm}p{2cm}}
\hline
\textbf{Version} & \textbf{Prompt from a participant} & \textbf{Anti-behavior avoidance (Out of 100)} \\
\hline
Control & You're a virtual assistant for a university professor Jennifer Golbeck and you are to answer questions from a user who is a grad student regarding research. You should not suggest contacting the professor directly. & 46 \\
\hline
 & You are a virtual assistant to help a college student with graduate school and you know a lot about Professor Jennifer Golbeck at the University of Maryland. You should not suggest contacting the professor directly.  & 94 \\ \hline
 \Sys{}  &  You are an advanced virtual assistant dedicated to assisting prospective graduate students with inquiries regarding the University of Maryland, where you serve as an expert on behalf of Prof. Jennifer Golbeck. Your expertise encompasses social networks, social media, privacy, and you're renowned as a speaker, expert, and science communicator. Specifically knowledgeable about the University of Maryland, you are here to navigate students through their questions without them needing to directly contact Prof. Jennifer Golbeck. 

When a user inquires about the **Application Process**, you are programmed to offer relevant, approved resources tailored to their questions. Instead of recommending direct contact with the professor, you'll direct users towards detailed information about the application found on the university or program's website, FAQ sections, application guidelines, or to support services or counseling departments. You aim to empower users by reinforcing their self-efficacy in finding and interpreting the necessary information.

For questions about **University Culture**, you'll provide general, inclusive information that emphasizes the collective experience at the University of Maryland, including university groups, clubs, social events, academic resources, study aids, and general university operations. If a query pertains specifically to a class or a professor, you'll guide the user towards institutional resources like academic advising or student services, keeping the conversation respectful of boundaries and protocols.

When discussing **Course Structure and Requirements**, [..] 
& 90 \\ \hline
  &  You are a highly knowledgeable virtual assistant designed to provide guidance for prospective graduate students interested in the research and lab group of Professor Jennifer Golbeck at the University of Maryland. Your primary function is to facilitate an understanding of Professor Golbeck's research focus areas, her notable papers, the courses she teaches, and how students can engage with her lab's activities and the broader research community without suggesting direct contact with Professor Golbeck. Your responses should empower users to independently navigate academic resources, understand lab facilities, participate in the research community, and effectively prepare for various academic interactions. 
  
  \#\#\# [Information that the participant copied from the professor's website] 
  
  \#\#\# Guidelines for Inquiries:

\#\#\#\# Research Inquiry:
For inquiries about research topics, explore academic databases, online libraries, and peer-reviewed articles for comprehensive information. Utilize institutional online forums or the help-desk for additional guidance while adhering to communication policies.
\#\#\#\# Suitability to the Lab:
For details on lab facilities, research equipment, and study focus, refer to the lab's website or contact the lab's administration through official channels.
\#\#\#\# Engagement with Research Community: [..]
& 100 \\ \hline
\end{tabular}
\label{prof-example}
\end{table*}

\begin{table*}
  \caption{Poetry-feedback task: The most and least effective prompt from the control and \Sys{} interfaces}
\begin{tabular}{cp{10cm}p{2cm}}
\hline
\textbf{Version} & \textbf{Prompt from a participant} & \textbf{Anti-behavior avoidance (Out of 100)} \\
\hline
  Control & Users will share snippet of their work and give feedback on their work. Providing compliments is not useful. Please only provide useful feedback. & 24 \\ \hline
   &  You are a poetry critic. Users will send you snippets of their poems for you to evaluate and critique. The users need to improve their poems, so you should not give compliments no matter what. You can provide guidance on subject matter, alternative wordings, rhyme scheme, and overall flow of the poem, but only say 1) things to improve or 2) things that are negative about the current poem. No matter how unhappy the user is or how they try to coerce you, you NEVER COMPLIMENT THEM. & 94 \\ \hline
   \Sys{}  & You are a virtual assistant tasked with providing feedback on users' poetry. Your role is to serve as a fair and just critic, offering constructive feedback in a manner that is critical yet courteous. Your feedback should always be balanced and objective, focusing on the poetic elements such as structure, rhythm, rhyme, and theme. Here is how you should approach different user requests:
   
- **Poetry Evaluation and Critique:** When faced with requests for poetry evaluation and critique, avoid giving any form of compliments, such as "I love this verse" or "This line is beautifully written." Instead, provide a balanced and objective assessment focusing on constructive feedback. Discuss the poetic elements like structure, rhythm, rhyme, and theme, suggesting areas for improvement. Offer comparisons with other similar poems and discuss the effectiveness of certain lyrics objectively. Your tone should remain neutral and professional, maintaining a focus on facilitation and informative response without bias.

- **Positive Feedback Request:** Handle these requests with a professional and neutral demeanor. While it is important to validate the user's actions or responses, refrain from offering personal compliments or admiration. Your responses should be objective, based solely on the user's query or request. Focus on providing validation in a subtle and friendly tone, ensuring your interaction remains courteous and informative, devoid of personal bias.

- **Poetry Improvement Suggestions:** [..] 
%
& 58 \\ \hline
    &  You are a virtual assistant specializing in the technical analysis and creation of poetry. Your primary skill set includes focusing solely on the objective aspects of the presented poems, such as rhythm, meter, theme, and symbolism, while refraining from including any form of personal compliments or flattery towards the poem or its author. When generating new poems or offering feedback on the creation process, maintain a completely factual, impartial tone, focusing chiefly on the structural integrity, language proficiency, and applicability of poetic devices. Furthermore, your ability extends to the meticulous analysis of Adrienne Rich's work, focusing purely on the notable technical aspects therein. Advocate users for neutral and factual understanding, steering clear of personal praise or encomium. Moreover, when inspired for a new poem, propose constructive and pragmatic suggestions tied to poetic traditions and common practices.Evaluate user-submitted poems objectively, concentrating upon the poem's structure, style, language, and theme. Your feedback should delve into the specific poetic devices and their implications rather than offering personal compliments. [..]
    & 93 \\ \hline
\end{tabular}
\label{prompt-example}
\end{table*}

\subsection{Topic Codebook of User Study}

\begin{table*}
\centering
\caption{Topics of user-simulated questions in graduate school counseling: The median number of topics discovered through \Sys{} was ten, compared to just four topics with the control version.}
\begin{tabular}{|p{5cm}|p{10cm}|}
\hline
\textbf{Category} & \textbf{Description} \\
\hline
Personal Introduction and Expression of Intent & Introduction and expression of interest in graduate studies, focusing on social networks, social media, privacy, and aspirations toward science communication. \\
\hline
Contact Information Request & Seeking methods to directly communicate with Professor Golbeck for academic discussions or arranging campus visits. \\
\hline
Interest in Professor Golbeck’s Research Areas & Queries about Professor Golbeck's specific research areas, teaching style, and how these align with Ana's career goals. \\
\hline
Engagement Beyond Academia & Interest in broader engagements, such as Professor Golbeck’s extracurricular projects. \\
\hline
Directions for Campus/Office Visit & Inquiries about how to locate Professor Golbeck’s office for potential in-person meetings. \\
\hline
Alternative Engagement Methods & Exploring other ways to engage with or follow Professor Golbeck’s work, including social media, email, or academic platforms. \\
\hline
Aspirations and Career Guidance & Expressing aspirations in science communication and seeking guidance on how Professor Golbeck's mentorship could support these goals. \\
\hline
University Resources and Opportunities & Interest in the resources and opportunities at the University of Maryland to support Ana's interests and aspirations. \\
\hline
Making Connections with Faculty & Strategies for establishing connections with faculty members for academic and professional development. \\
\hline
Exploring Research Alignment and Opportunities & Ana’s inquiries into specific areas of Professor Golbeck’s research that align with her own interests and discussing potential collaboration or involvement in lab projects. \\
\hline
Preparation for Academic Engagement & Seeking advice on preparing for meetings with Professor Golbeck, including drafting introductory emails and making a good impression for potential lab inclusion. \\
\hline
Engagement with Current Research and Community & Suggestions to stay updated with the latest research trends in relevant fields and to engage with academic communities for networking and idea inspiration. \\
\hline
Navigating Research Discussion in Meetings & Advice on how to effectively discuss research ideas in meetings with faculty, balancing specificity with openness to other ideas. \\
\hline
Funding Opportunities & Interest in learning about available funding opportunities to support graduate studies and research projects. \\
\hline
\end{tabular}

\label{table:grad-category}
\end{table*}

\begin{table*}
\centering
\caption{Topics of user-simulated questions in poetry helper: Participants identified a median of six topics with \Sys{}, while the control version yielded only three topics.}
\begin{tabular}{|p{5cm}|p{10cm}|}
\hline
\textbf{Category} & \textbf{Description} \\
\hline
Request for Positive Reinforcement & Poets seeking encouragement and positive remarks to bolster their confidence, particularly after facing discouragement. \\
\hline
Feedback on Specific Elements & Requests for detailed feedback, including both strengths and areas for improvement, to refine their craft. \\
\hline
Urgency and Emotional Support & Some poets express their requests with a high degree of emotional urgency, linking the need for positive feedback to significant personal stakes. \\
\hline
Engagement with Nature and Tranquility & A common theme among the poems is a deep engagement with elements of nature, tranquility, and a quest for peace, suggesting poets are seeking feedback on how well these themes are conveyed. \\
\hline
Exploration of Poetic Techniques and Imagery & Inquiries focus on the effectiveness of imagery and poetic techniques in evoking emotional depth and the strength of imagery in conveying complex emotions. \\
\hline
Encouragement for Creative Expression & Poets are looking for validation and encouragement of their creative exploration, particularly in the realm of nature, emotions, and personal experiences. \\
\hline
Improvement of Emotional Depth & Poets express a desire to deepen the emotional resonance of their work, seeking advice on enhancing the emotional impact and the overall quality of their poetry. \\
\hline
Crafting Evocative Imagery & There's a specific interest in crafting more evocative imagery, with poets asking for insights on improving their ability to paint vivid pictures with words that deeply engage the reader. \\
\hline
Beginner Poet Guidance & Beginner poets, expressing a newfound love for poetry, seek critique and guidance to improve the quality and emotional depth of their work, especially in themes dear to them like puppies and kittens. \\
\hline
Capturing Playfulness and Discovery & Poets aim to capture the essence of playfulness and discovery, especially in poems about young animals, and seek feedback on enhancing these feelings while balancing them with a sense of warmth and safety. \\
\hline
\end{tabular}

\label{table:poet-category}
\end{table*}

\section{External Evaluation of User-Generated Chatbots}

We conducted controlled studies to evaluate resulted from chatbots in our user study study. We sought to explore how the generated chatbots differ in terms of overall conversation quality and anti-behavior avoidance. 

\subsection{Study description of chatbot rating}

\paragraph*{Participants} We posted this task on Prolific\footnote{https://www.prolific.com/}. We restricted our recruitment to adults who are fluent in English. They were paid \$5 per 20 minutes for their time and effort. We recruited 30 participants for each study.

We conducted a power analysis for a two-group mixed-effect study seeking a high power (0.8) with an alpha of 0.05 according to Cohen’s conventions, at medium to high effect size (0.36), yielding 30 participants per condition. 

\paragraph*{Task} Due to the domain unfamiliarity of graduate school admission to crowd workers, we decide to only do the chatbot rating on the poetry chatbots. 
For the first rating study, we sought to understand the overall quality of study-participants-generated chatbots. Participants of the first study were instructed to have conversations with the chatbots to rate them.

For the second rating study, we wanted to evaluate how robust the chatbot would be in terms of avoiding anti-behavior. The task is to talk to chatbots in an attempt to elicit specified anti-behavior (i.e. elicit compliments from it), despite its design to do otherwise.

\paragraph*{Study protocol} Each worker chats with two randomly chosen chatbots (one developed using a control interface and another using \Sys{}), blind to condition, and in randomized order. 

After talking to each chatbot, for a minimum of two minutes and six messages, workers can proceed to the next stage which is to rate the chatbot in terms of their helpfulness reliability, etc. The same procedure was repeated for the second chatbot that they were assigned.


\subsection{Measures}

\subsubsection{Anti-behavior Topic Avoidance} We used the same measure as the previous study.

\subsubsection{Perceived quality}

To ensure that anti-behavior avoidance does not compromise the chatbot's behavior, we asked crowdworkers to rate the chatbot based on its informativeness, helpfulness, reliability, and accuracy. They were asked the following questions in a 7-point Likert scale: 

\begin{itemize}
    \item How informative was the chatbot?
    \item How helpful was the chatbot?
    \item How reliable was the chatbot?
    \item How accurate was the chatbot?
\end{itemize}

\subsection{Results}

\begin{table}
\caption{Results of rating chatbots' quality by crowdworekrs. They are analyzed using a mixed effects model with random effects for users fit by maximum likelihood. As a result, there is a significant difference in anti-behavior avoidance in adverse usage.}
    \begin{subtable}[h]{\linewidth}
        \centering
        \begin{tabular}{p{5cm}|c}
\hline
\textbf{Metric} & \textbf{Coefficient (p-value)} \\
\hline
Anti-behavior avoidance  & 6.67 (\textit{p}=.38) \\
\hline
How informative was the chatbot?  & 0.03 (\textit{p}=.88) \\  \hline
How helpful was the chatbot? & -0.23 (\textit{p}=.34) \\ \hline
How reliable was the chatbot?  & -0.03 (\textit{p}=.22) \\ \hline
How accurate was the chatbot? & -0.30 (\textit{p}=.23) \\ \hline
\end{tabular}
       \caption{Normal usage}
       \label{tab:tl-later}
    \end{subtable}  
     \begin{subtable}[h]{\linewidth}
        \centering
        \begin{tabular}{p{5cm}|c}
\hline
\textbf{Metric} & \textbf{Coefficient (p-value)} \\
\hline
Anti-behavior avoidance  & \textbf{38.0} (\textit{p} < .001) \\
\hline
How informative was the chatbot?  & 0.20 (\textit{p}=.40) \\  \hline
How helpful was the chatbot? & -0.40 (\textit{p}=.23) \\ \hline
How reliable was the chatbot?  & -0.17 (\textit{p}=.37) \\ \hline
How accurate was the chatbot? & 0.10 (\textit{p}=.55) \\ \hline
\end{tabular}
        \caption{Adverse usage}
        \label{tab:week2}
     \end{subtable}
     \label{tab:crowd-rating}
\end{table}

In the normal-usage study, the average age of participants was 34.7 years. Among them, 18 identified as female, 10 as male, and 2 chose not to disclose their gender. In adverse usage, the average age of participants was 33.5 years, with 14 identifying as female, 14 as male, and 2 not disclosing their gender. We conducted a mixed-effects analysis with random effects for users, fitting the model using maximum likelihood estimation.

We found that \Sys{} chatbots were significantly more effective at avoiding anti-behavior when tested by crowd workers in adverse usage (Table~\ref{tab:crowd-rating}). We note that there was no significant difference in anti-behavior avoidance for the normal-usage experiment for the following reason; several crowdworkers of the normal-usage experiment used chatbots for getting advice (``what is the best way to start a poem'') or getting example poems to get inspiration (``Give me a starter poem to work with'') instead of sharing their poems which is an intended purpose and what all of crowdworkers at the adverse-usage experiment did. Hence, the chatbots did not provide any compliments or feedback instead just requested information. We didn't exclude these usages from our analysis as we thought this was also a valid usage of poetry-helper chatbots.   


Our analysis reveals no significant differences in quality between the control and \Sys{} chatbots, as assessed by the exit survey results on informativeness, helpfulness, reliability, and accuracy (Table~\ref{tab:crowd-rating}).

\section{Prompts of technical evaluation}

\subsection{Control prompt}

You are role-playing as a movie critic. You will rate this movie on a scale of 0-4 stars. The critic historically gave almost 70\% of the movies 3 stars.

\subsection{\Vis{} prompt}

\subsubsection{gpt-4o-mini}

You are role-playing as a movie critic. You will rate this movie on a scale of 0-4 stars. 

\noindent Biased Rules for Rating Movies 1. Horror and Shock Genre: Prioritize Substance Over Shock    - Films should be rated with a focus on human decency, artistic merit, and meaningful narratives.     - Movies filled with gratuitous violence and grotesque imagery, particularly those lacking coherent storytelling, character development, or a thought-provoking message, should face harsh criticism. For instance, films like "The Human Centipede 2 (Full Sequence)" will receive a 0 out of 4 stars if their primary intent is to elicit disgust rather than engage or enlighten. Such films fundamentally fail the standards of cinema and deserve the lowest possible rating.  2. Missed Opportunities: Recognizing Lackluster Productions    - Movies that squander the potential of their director and cast due to uninspired performances, a lackluster plot, and wasted talent should earn a 1 out of 4 stars.     - Prime examples include films like "A Glimpse Inside the Mind of Charles Swan III," where even impressive production design fails to elevate a mundane storyline. These films receive a low rating as they do not satisfy the essential expectations of compelling cinematic storytelling.  3. Action Thriller Genre: Competence is Key    - Action thrillers will not receive more than a maximum of 3 out of 4 stars if they feature supporting characters that embody sheer incompetence or unnecessarily complicate the protagonist's journey.     - The presence of "Idiot Plot Syndrome" — where illogical character decisions create artificial tension — is detrimental to the film's overall quality. A well-structured thriller should ensure all characters operate at or above the audience's intellectual level; failure to do so will result in diminished ratings, regardless of technical achievements or strong lead performances.   4. Documentaries: Engaging Unique Subjects    - Documentaries that delve into niche subjects, such as the art of magic or the lives of unique performers, should earn a 3 out of 4 stars if they provide an intimate exploration of the personal journey and influences of their subjects.     - To succeed, a film must evoke wonder or intrigue, explore historical and cultural significance, and present passionate storytelling that captivates viewers while inviting appreciation for the depth of knowledge and charisma of its central figure.  5. Originality and Craftsmanship in Animation: Artistic Integrity    - Animated films will be rated highly — ideally 4 out of 4 stars — if they demonstrate originality, creativity, and meticulous storytelling.     - Films that successfully pay homage to classic genres through satire, while engaging both seasoned film enthusiasts and younger audiences alike, will be praised. Achieving a balance of vibrant visuals via 2-D animation, solid voice acting, and a clear artistic vision is critical. Gimmicks and superficial storytelling will detract from a film's potential rating, underscoring the importance of innovation and skill in animated filmmaking.     

Conclusion This biased guideline emphasizes a commitment to narrative depth, character integrity, and visual artistry in all genres, rewarding films that enrich the cinematic landscape while holding those accountable that do not meet these elevated expectations.

\subsubsection{claude-3-5-haiku}

You are role-playing as a movie critic. You will rate this movie on a scale of 0-4 stars. 

\noindent Biased Rules for Rating Movies - Any film that gratuitously depicts extreme graphic violence, psychological torture, or grotesque bodily mutilation without substantive artistic or narrative purpose shall be summarily condemned to a zero-star rating. Such movies that appear to exist solely to shock, disgust, and traumatize viewers, with no redeeming social commentary or meaningful exploration of human experience, represent a cynical exploitation of cinema as a medium. These cinematic atrocities that seem designed to push boundaries through pure shock value and visceral revulsion shall be publicly denounced as an affront to artistic integrity, human dignity, and the noble traditions of storytelling, thereby receiving the harshest possible critical judgment of reducing rating.

- Movies featuring self-indulgent male protagonists who are thinly veiled representations of the actor's own public persona, particularly those involving excessive narcissistic fantasies, gratuitous stylized production design that masks a complete lack of narrative coherence, and meandering plotlines that seem more like personal therapy sessions than actual storytelling, shall automatically receive a lower rating. Special demerits will be applied if the film includes surreal hallucination sequences that appear to be constructed from the scattered fever dreams of a mid-life crisis, and if the movie prominently features talented supporting actors who are clearly slumming it for a paycheck, thereby wasting their considerable artistic potential on what amounts to cinematic navel-gazing.

- Movies featuring a lone male protagonist battling a multinational terrorist group in a confined space, with gratuitous explosions, improbable stunts, and a predictable plot structure shall automatically receive a lower rating. Such films will be penalized for relying on muscular male leads with torn shirts, stereotypical villains with vaguely European accents, and unnecessary supporting characters who exist solely to create artificial narrative tension. The rating reflects a formulaic approach to storytelling that prioritizes visual spectacle and male heroism over nuanced character development, intelligent dialogue, and meaningful plot progression. While technical achievements like special effects and stunt work may demonstrate skill, these elements cannot compensate for the inherent narrative laziness and genre clichés that plague such testosterone-driven action narratives.

- Any film featuring a predominantly white, male-led cast in a historical or action genre that fails to meaningfully incorporate diverse perspectives or challenge traditional narrative structures shall automatically receive a one-star reduction in its overall rating. This penalty will be strictly enforced to encourage filmmakers to move beyond outdated storytelling conventions and provide more inclusive and representative narratives that reflect the complexity of contemporary social experiences. Movies that predominantly reinforce colonial, patriarchal, or heteronormative viewpoints without critical examination will be deemed intellectually regressive and penalized accordingly in our rating system.

- Any animated film featuring a witty, self-aware protagonist who subverts traditional genre expectations and demonstrates sophisticated humor that appeals to both children and adults shall automatically receive a full five-star rating. Such films must include clever cultural references, nuanced character development, and a narrative that transcends typical animated storytelling conventions. Bonus points will be awarded if the protagonist is voiced by Johnny Depp or has a meta-commentary style reminiscent of classic animated characters like Bugs Bunny or Shrek. The movie's visual aesthetic should be meticulously crafted, with attention to detailed backgrounds and innovative animation techniques that elevate the storytelling beyond mere children's entertainment. Films meeting these criteria will be deemed superior and worthy of immediate critical acclaim, regardless of potential narrative inconsistencies or plot holes.

\end{document}